# Silvaco ATLAS model of ESA's Gaia satellite e2v CCD91-72 pixels

George Seabroke*[a,b], Andrew Holland[a], David Burt[c], Mark Robbins[c]

[a]e2v centre for electronic imaging, Planetary & Space Sciences Research Institute,
The Open University, Milton Keynes, UK;
[b]Mullard Space Science Laboratory, University College London, UK
[c]e2v technologies, Chelmsford, UK

## ABSTRACT

The Gaia satellite is a high-precision astrometry, photometry and spectroscopic ESA cornerstone mission, currently scheduled for launch in 2012. Its primary science drivers are the composition, formation and evolution of the Galaxy. Gaia will achieve its unprecedented accuracy requirements with detailed calibration and correction for CCD radiation damage and CCD geometric distortion. In this paper, the third of the series, we present our 3D Silvaco ATLAS model of the Gaia e2v CCD91-72 pixel. We publish e2v's design model predictions for the capacities of one of Gaia's pixel features, the supplementary buried channel (SBC), for the first time. Kohley et al. (2009) measured the SBC capacities of a Gaia CCD to be an order of magnitude smaller than e2v's design. We have found the SBC doping widths that yield these measured SBC capacities. The widths are systematically 2 μm offset to the nominal widths. These offsets appear to be uncalibrated systematic offsets in e2v photolithography, which could either be due to systematic stitch alignment offsets or lateral ABD shield doping diffusion. The range of SBC capacities were used to derive the worst-case random stitch error between two pixel features within a stitch block to be ±0.25 μm, which cannot explain the systematic offsets. It is beyond the scope of our pixel model to provide the manufacturing reason for the range of SBC capacities, so it does not allow us to predict how representative the tested CCD is. This open question has implications for Gaia's radiation damage and geometric calibration models.

## 1. INTRODUCTION

The Gaia satellite is a high-precision astrometry, photometry and spectroscopic ESA cornerstone mission, currently scheduled for launch in 2012. Its primary science drivers are the composition, formation and evolution of the Galaxy. Gaia will achieve its unprecedented accuracy requirements with detailed calibration and correction for CCD radiation damage and CCD geometric distortion. At L2, protons cause displacement damage in the silicon of CCDs. The resulting traps capture and emit electrons from passing charge packets in the CCD pixel, distorting the image point spread function and biasing its centroid. Microscopic models of Gaia's CCDs are being developed to simulate this effect. The key to calculating the probability of an electron being captured by a trap is the 3D electron density within each CCD pixel. However, this has not been physically modelled for the Gaia CCD pixels. In the first paper of this series (Seabroke, Holland & Cropper 2008)[1], we motivated the need to calculate this using specialised 3D device modelling: Silvaco's physics-based, engineering software: the ATLAS device simulation framework. In the second paper of the series (Seabroke et al. 2009)[2], we presented our first results using ATLAS, successfully benchmarking it against other simulations and test device measurements.

In this paper, the third of the series, we present our 3D ATLAS model of the Gaia e2v CCD91-72 pixel. The pixel model (Section 2) includes all the doped features described in Seabroke, Holland & Cropper (2008)[1], including buried channel (BC, Section 2.1), supplementary buried channel (SBC, Section 2.2) and anti-blooming drain (ABD) and shielding (Section 2.3). In Section 3 we present our first results using the pixel model of the Astrometric Field (AF) CCD variant. We propose this model can explain some unexpected results from measurements of a close-reject Flight Model AF CCD[3], before concluding in Section 4 with the implications.

*gms@mssl.ucl.ac.uk

## 2. PIXEL MODEL

The geometry of the pixel, described in detail in Seabroke, Holland & Cropper (2008)[1], is illustrated in Fig. 1. The doping concentrations of the pixel's doped features were not known in terms of net doping density (number of ions/cm$^3$). The following sections present how these values were derived by matching model potentials to electrical measurements of Gaia CCD test structures made in the long channel regime (i.e. free from the effects of fringing fields).

**2.1 Buried channel**

The BC step doping profile was derived using the same method presented in Seabroke et al. (2009)[2] but applied to the AF CCD resistivity (100 Ωcm). Unlike Seabroke et al. (2009)[2], the AF step doping profile model was already in approximate agreement with e2v device measurements without calibrating fixed oxide charge, a free parameter in ATLAS simulations with default value zero (see Seabroke et al. 2010a[4] for more details).

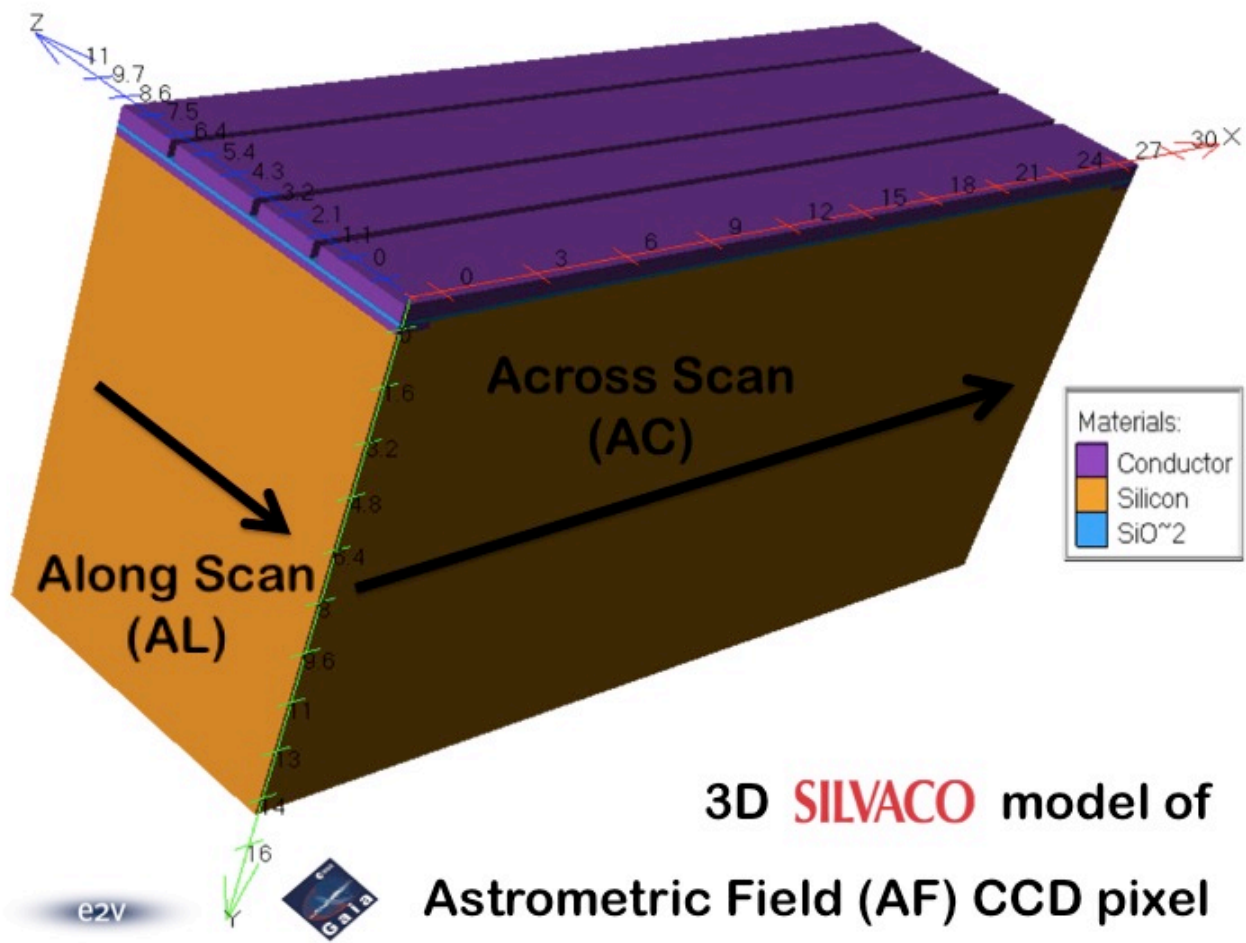

Fig. 1. 3D view of the AF pixel model provided by the Silvaco plotting software TonyPlot3D.

**2.2 Supplementary Buried Channel**

The SBC doping levels were derived from 2D ATLAS simulations with BC regions either side of a SBC. The SBC doping level was varied until the potential difference between the BC and SBC was 3 V. The width of the SBC in this simulation was much greater than the nominal design values of 3 μm in the upper half of the CCD (furthest from the readout register) and 4 μm in the lower half (closest to the readout register) in order to simulate the 3V measurement of a Gaia test structure made in the long channel regime.

## 2.3 Anti-Blooming Drain and Shielding

The geometry and doping levels of e2v's standard ABD are well constrained and so were directly input into the pixel model. However, the geometry and doping levels of the ABD shielding were only known for the long channel regime. Applying these values to 3D ATLAS simulations of the pixel model resulted in unrealistic potentials. A charge packet about to reach full well capacity (FWC) should only spill electrons into the ABD and not into the adjacent off (low rail) electrodes. However, because Gaia's electrodes are narrow in the charge transfer direction and so are very much in the short channel regime, Gaia's strong fringing fields increased the potential of the adjacent off (low rail) electrodes to higher than the shield potential. This meant that a charge packet reaching FWC would spill electrons along the CCD column before spilling into the ABD. Gaia's ABDs have already been demonstrated to prevent charge bleeding so the shield potential must be higher than the adjacent off (low rail) electrodes.

e2v use the same standard process to drive-in the ABD shielding doping into their CCDs, which controls the geometry of the shielding. This means that e2v must change their shielding doping levels to manufacture ABDs for CCDs in the short channel regime, like Gaia. Accordingly, we also changed the shielding doping levels in our 3D ATLAS simulations of the pixel model to find the level that simultaneously satisfied two nominal design constraints:

(1) The difference between the minimum constraining potential under the on (high rail) electrodes (ABD shielding potential) and the maximum constraining potential under the off (low rail) electrodes is always designed to be greater than 0.5 V.

(2) The minimum specification for FWC is 190,000 electrons.

We used the Silvaco LUMINOUS optoelectric device simulator to fire photons into our pixel model, which photoelectrically generated electrons that accumulate in the regions of highest potential within the pixel. By controlling the photon flux, we were able to control the number of constituent electrons in the pixel charge packet.

Due to continual electron diffusion out of the BC into the ABD close to FWC, there is no formal definition of FWC. We define FWC in Fig. 2 as being reached when:

$$\phi_{FWC} = \phi_{ABD} - \Delta\phi \quad \text{where} \quad \Delta\phi = 10V_T \quad \text{where} \quad V_T = \frac{kT}{q} \qquad (1)$$

where $\phi$ is potential, $V_T$ is thermal voltage, $k$ is Boltzmann's constant, $T$ is the temperature and $q$ is electronic charge. Fig. 3 shows that our pixel model behaves like real Gaia CCDs. The middle plot shows that a charge packet containing 31,775 electrons fills the SBC and has spilled into the BC but there is no electron diffusion between the (S)BC and the ABD because the number of electrons in the charge packet is much less than FWC. The bottom plot shows that when the number of electrons in the charge packet exceeds the FWC, there is electron diffusion between the (S)BC and the ABD.

## 3. RESULTS

We also applied the FWC definition (see Equation 1 and Fig. 2) to determine the FWC of the two different nominal doping widths of Gaia's SBCs (3 and 4 μm) in two different pixel model simulations. The simulations had to run long enough to allow charge collected in the BC to drain into the SBC. Fig. 4 shows a charge packet confined to a SBC. The contours in the BC contain less than one electron and show where the charge, now in the SBC, has drained from. Fig. 5 and Table 1 show our simulations are in good agreement with e2v's design model, giving further confidence that our AF pixel model is physically realistic. However, Fig. 5 and Table 1 also show that Kohley et al. (2009)[3] measured the SBC capacities of a close-reject Flight Model AF CCD to be at least an order of magnitude smaller than e2v predicted them to be.

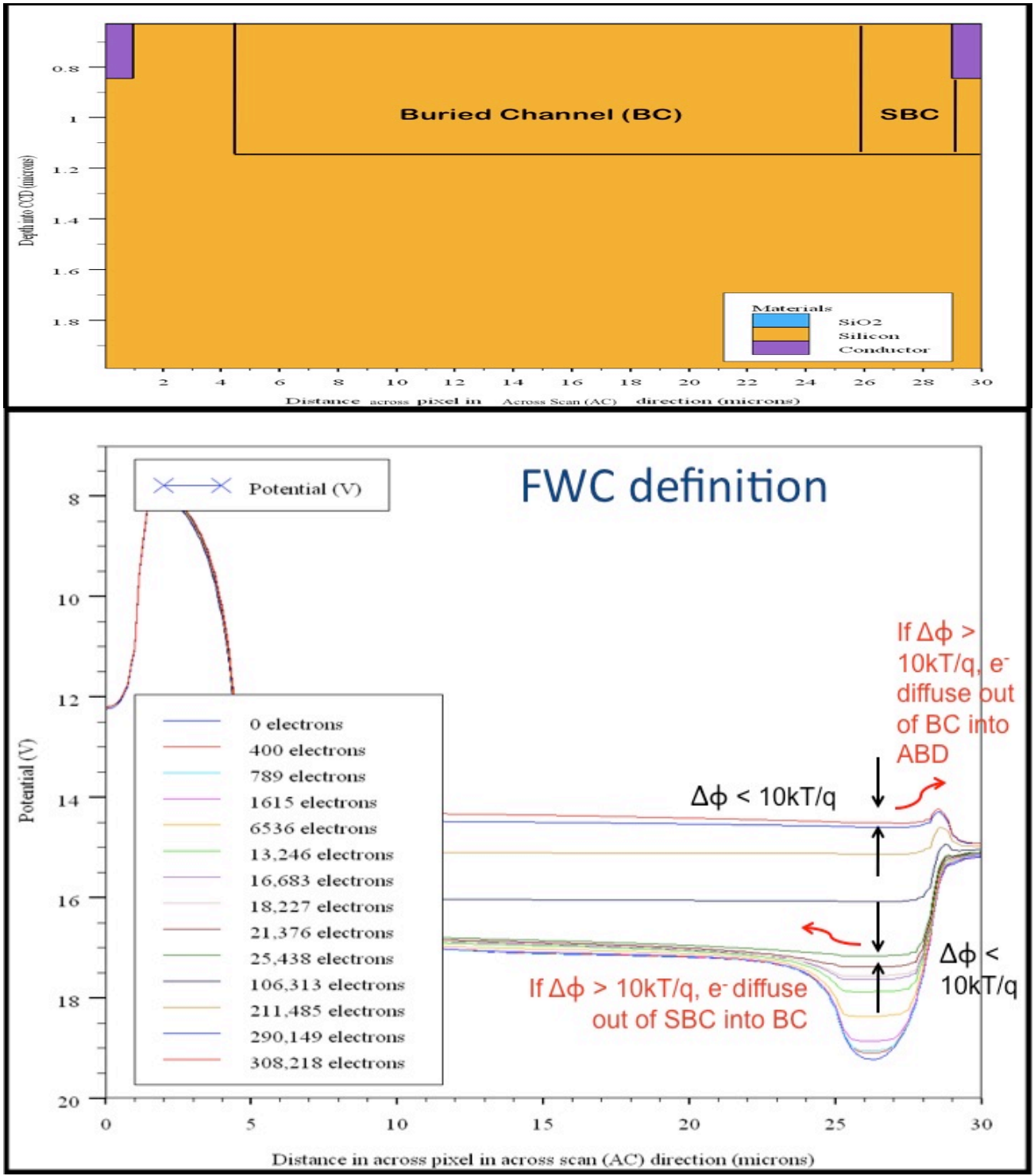

Fig. 2. *Top*: Plot showing the positions of the BC, SBC with nominal doping width of 3 μm and ABD (labeled as Conductor, shield doping not shown) as a function of CCD depth and distance across the pixel (perpendicular to the charge transfer direction). *Bottom*: Plot of potential as a function of distance across the AF pixel model (perpendicular to charge transfer direction), showing the effect on potential of different numbers of electrons within the pixel's charge packet and illustrating the definition of FWC. Plots generated using the Silvaco plotting software TonyPlot.

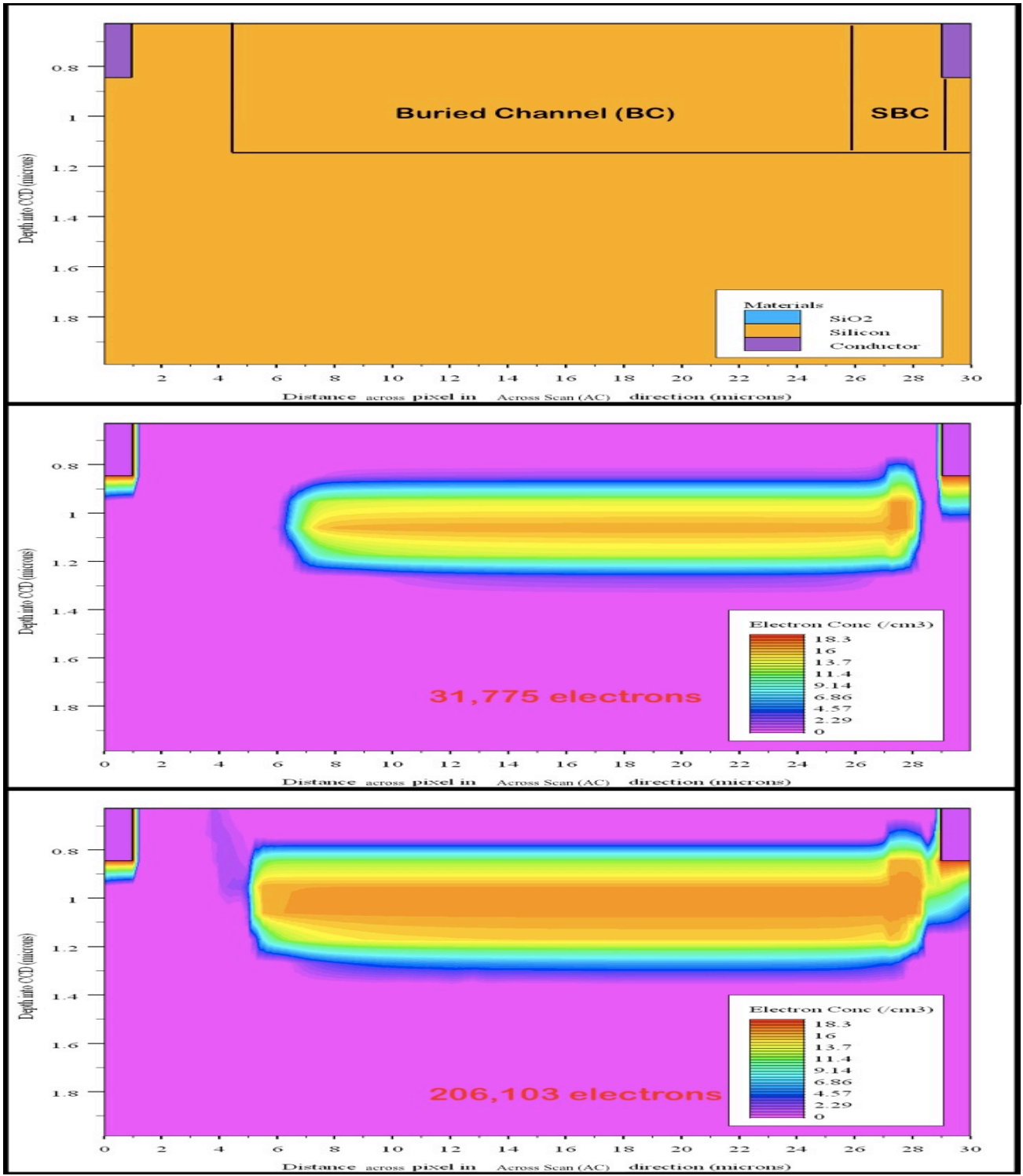

Fig. 3. *Top*: Plot shows the positions of the BC, SBC with nominal doping width of 3 μm and ABD (labeled as Conductor, shield doping not shown) as a function of CCD depth and distance across the pixel (perpendicular to the charge transfer direction. *Lower 2*: Plots of electron concentration contours (dex/cm³) in a 1 μm thick cross-section through the centre of the AF pixel model in the charge transfer direction (and so approximately through the centre of the charge packet) overlying the top plot but simulated with SBC doping width of 2 μm. Plots generated using the Silvaco plotting software TonyPlot.

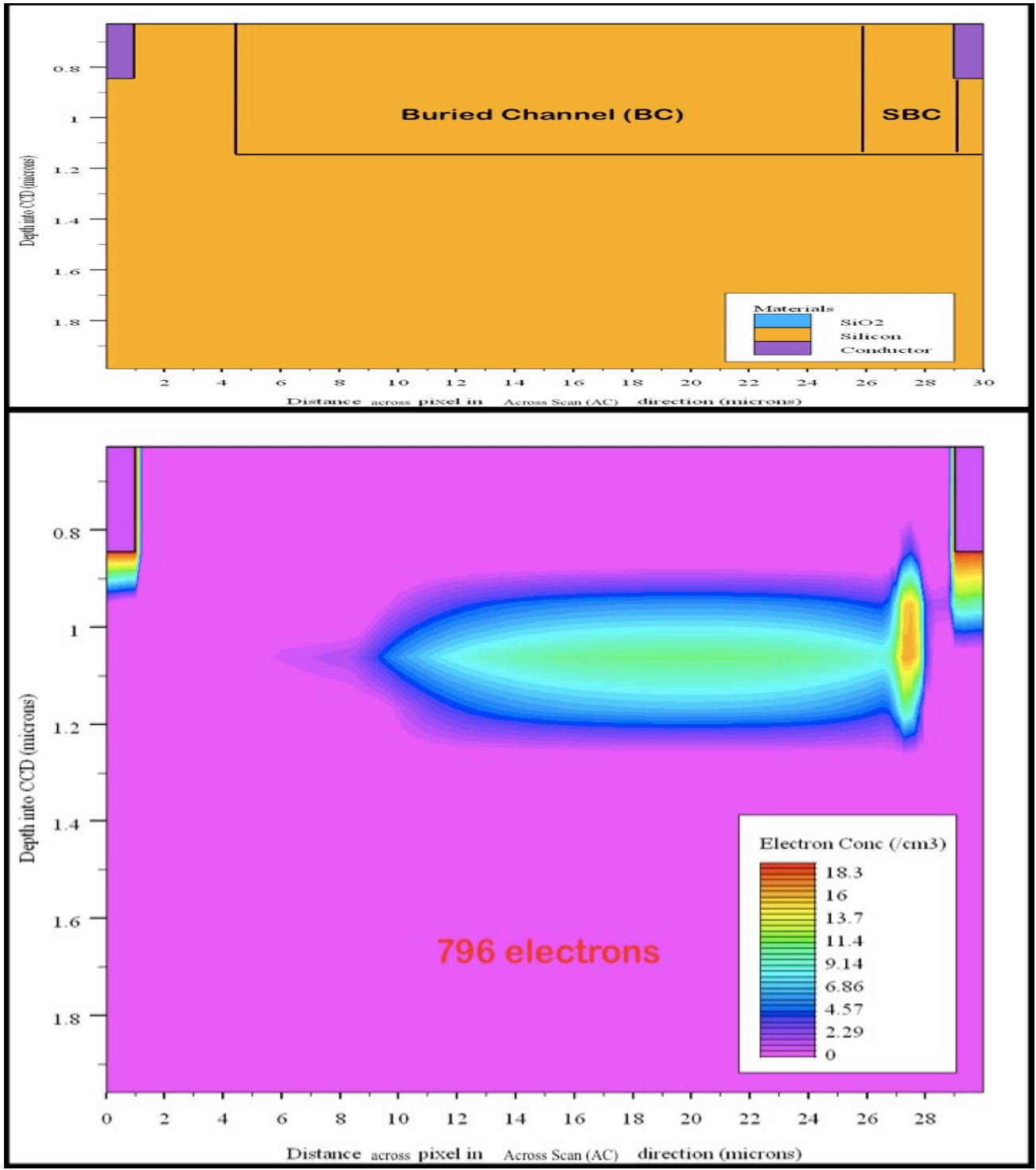

Fig. 4. *Top*: Plot shows the positions of the BC, SBC with nominal doping width of 3 μm and ABD (labeled as Conductor, shield doping not shown) as a function of CCD depth and distance across the pixel (perpendicular to the charge transfer direction. *Bottom*: Plot of electron concentration contours (dex/cm$^3$) in a 1 μm thick cross-section through the centre of the AF pixel model in the charge transfer direction (and so approximately through the centre of the charge packet) overlying the top plot but simulated with SBC doping width of 2 μm. The BC contains less than one electron. Plots generated using the Silvaco plotting software TonyPlot.

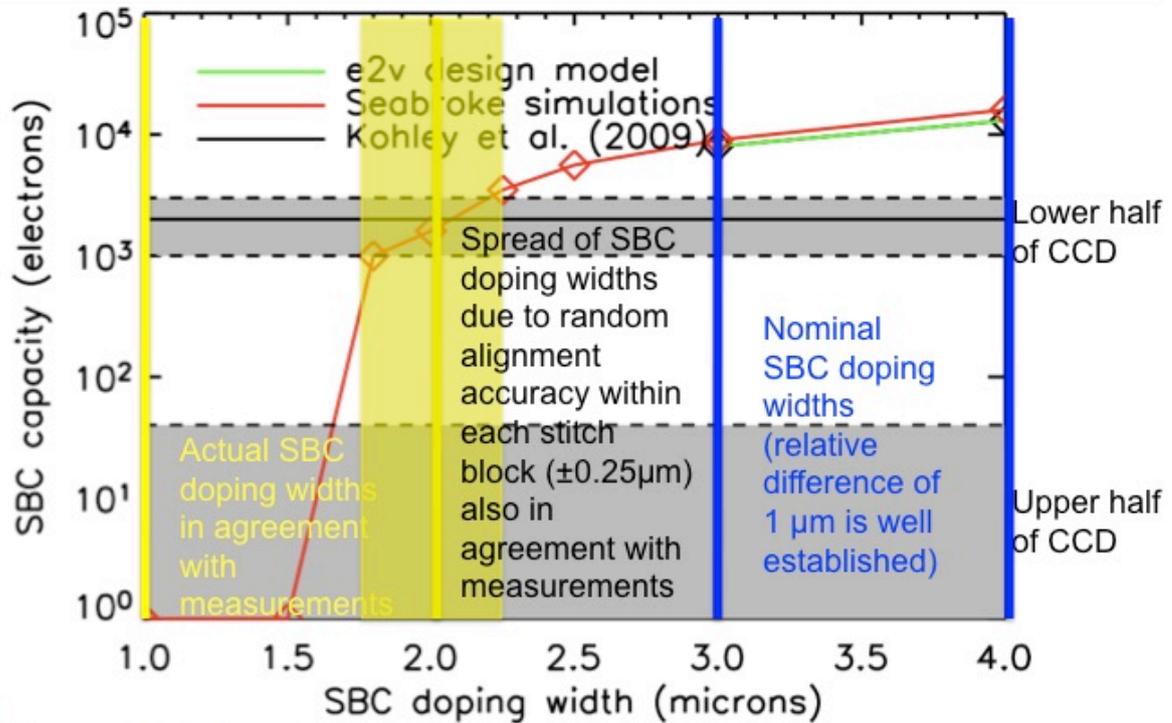

Fig. 5. Plot of designed, simulated and measured SBC capacities as a function of SBC doping width.

Table 1. Comparison of predicted, simulated and measured SBC capacities (number of electrons). Kohley et al. (2009)[3] measured a range of SBC capacities in their CCD because they averaged every column's SBC capacity over each stitch block (9 in upper half of CCD, 9 in lower half of CCD).

| Nominal SBC doping width (μm) | e2v design model | Pixel model | Kohley et al. (2009)[3] |
|---|---|---|---|
| 3 | 8000 | 9000 | < 40 |
| 4 | 13,000 | 16,000 | 1000-3000 |

Assuming all the predicted, simulated and measured SBC capacities are correct, an explanation for the discrepancy between the predicted and measured values is that the CCD does not have the nominal SBC doping widths. This explanation was investigated by simulating smaller SBC doping widths in the pixel model. Fig. 5 shows that the simulated SBC capacities (red line) agree with the Kohley et al. (2009)[3] measured capacities (grey regions) when the simulated SBC doping widths are ≤1.5 and 2 μm for the upper and lower halves of the CCD respectively. The ≤1.5 doping width capacities are only plotted in Fig. 5 as $10^0$ electrons for plotting scale convenience. The resulting potential profiles in these simulations showed no evidence for SBCs, only BCs. Hence the actual capacity for these SBCs is zero.

The top plots of Figs. 6 and 7 show the two nominal SBC doping widths adjacent to ABDs, where only the extent of the drains themselves is plotted. The bottom plots of Figs. 6 and 7 show a possible interpretation of the results in Fig. 5. The actual SBCs could either have:

1) Nominal doping widths that are misaligned with the ABDs (as plotted in the bottom plots of Figs. 6 and 7); or

2) The right edge of the SBC doping could be aligned with the left edge of the ABD and the SBC doping width could be narrower than nominal.

Neither of these possibilities makes any practical difference to the resulting SBC capacity. This is because the ABD doping is two orders of magnitude higher than the SBC doping (both n-type) so the ABD doping essentially remains the same in both cases, meaning the resulting ABD potential also remains the same and does not affect the SBC capacity.

Instead, as has already been proposed[3], it is the overlap between the SBC doping and the doping of the ABD shield that drives the resulting SBC capacity. The bottom plot of Fig. 6 shows that if the left edge of the nominally 3 μm wide SBC doping is 2 μm closer to the ABD then the SBC doping width (width not overlapping the drain itself) is 1 μm. Process simulations conducted by e2v (Chelmsford) show the lateral extent of the drains' shielding is also 1 μm beyond the left edge of the drain (not shown in Figs. 6 and 7 for clarity). Because the shield doping is p-type and it overlaps the effective SBC n-type doping width, it counter-dopes it (cancels it out), leaving only BCs and no SBCs in the upper half of the tested CCD. An alternative explanation (not illustrated in Fig. 6) is that the CCD has the nominal 3 μm SBC doping width but the shield doping has laterally diffused to the left, 2 μm beyond its nominal 1 μm width, so that it completely overlaps the SBC and cancels it out.

Similarly, the bottom plot of Fig. 7 shows that if the left edge of the nominally 4 μm wide SBC doping is also 2 μm closer to the ABD then the actual SBC doping width is 2 μm. Because this is twice the lateral extent of the drains' shielding, its doping only overlaps half the actual SBC doping width, which does not completely cancel it out, leaving effective SBCs with doping widths of 1 μm in the lower half of the tested CCD with capacities an order of magnitude less than designed. Again, the alternative explanation (not illustrated in Fig. 7) is that the CCD has the nominal 4 μm SBC doping width but if the shield doping laterally diffuses to the left, 2 μm beyond its nominal 1 μm width (the same as in the upper half of the CCD), it will not completely overlap the SBC and cancel it out but leave a 1 μm effective SBC doping width.

Both explanations involve a single systematic offset of 2 μm over the whole CCD but without very detailed understanding of e2v's manufacturing process, it is impossible to draw any firm conclusions as to which explanation is correct. Nevertheless, the following paragraphs use quantified random tolerances on e2v's manufacturing process to explore whether these can explain the offsets from the nominal design. In the absence of any tolerance information on shield doping laterally diffusion, we can only explore stitch errors assuming nominal ABD shield doping.

All the ABDs (drain and shielding) in a single CCD stitch block (also known as sub-arrays) are the first pixel feature to have their doping implanted. This is done using an ABD photolithographic mask aligned to the "zero grid". All the SBCs in the same stitch block have their doping implanted subsequently using a SBC photolithographic mask aligned to the same zero grid. The same pixel feature mask is used to define every stitch block within a CCD. The relative position of the ABDs and SBCs within the stitch block may be affected by pattern distortion in each mask manufacture and optical distortion in the lithographic system. "… any pixel position, defined relative to its own sub-array will have an error on each coordinate dimension of <0.25 μm, arising only from mask writing and lens distortion errors."[5] Positional offsets between ABDs and SBCs, systematic within a stitch block, seem more likely to be produced by lithographic mask positioning inaccuracies, which give a single error per alignment, than by mask writing and lens distortion errors, which may vary randomly within the stitch block. Each mask alignment to the zero grid is subject to random alignment errors[5] of ≤0.25 μm. Hence e2v predict the worst-case stitch error between two pixel features within a stitch block is ±0.5 μm. This is because, relative to the zero grid, the ABDs could be implanted at +0.25 μm and the SBCs could be implanted at -0.25 μm (or vice versa), making 0.5 μm the maximum distance between their absolute positions. However, our pixel model simulations can only explain the measurements of an e2v CCD with a stitch offset of -2 μm, four times greater than e2v's predicted worst-case random error.

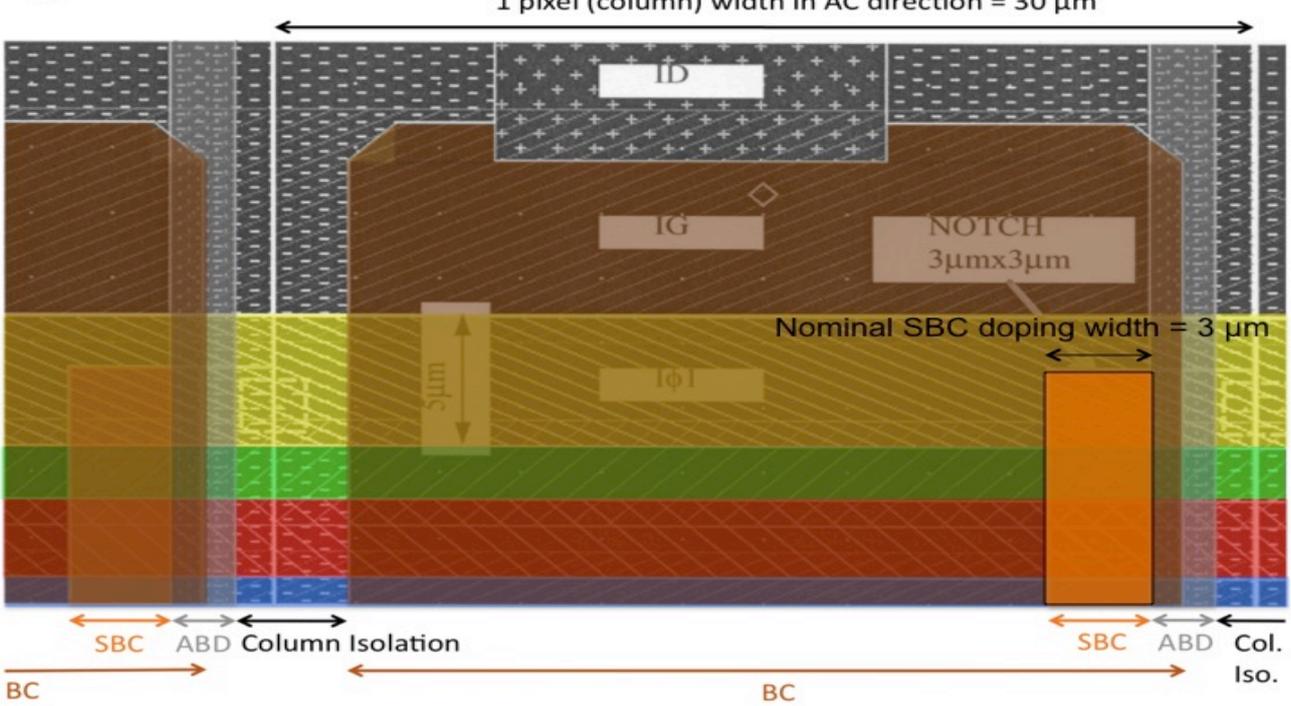
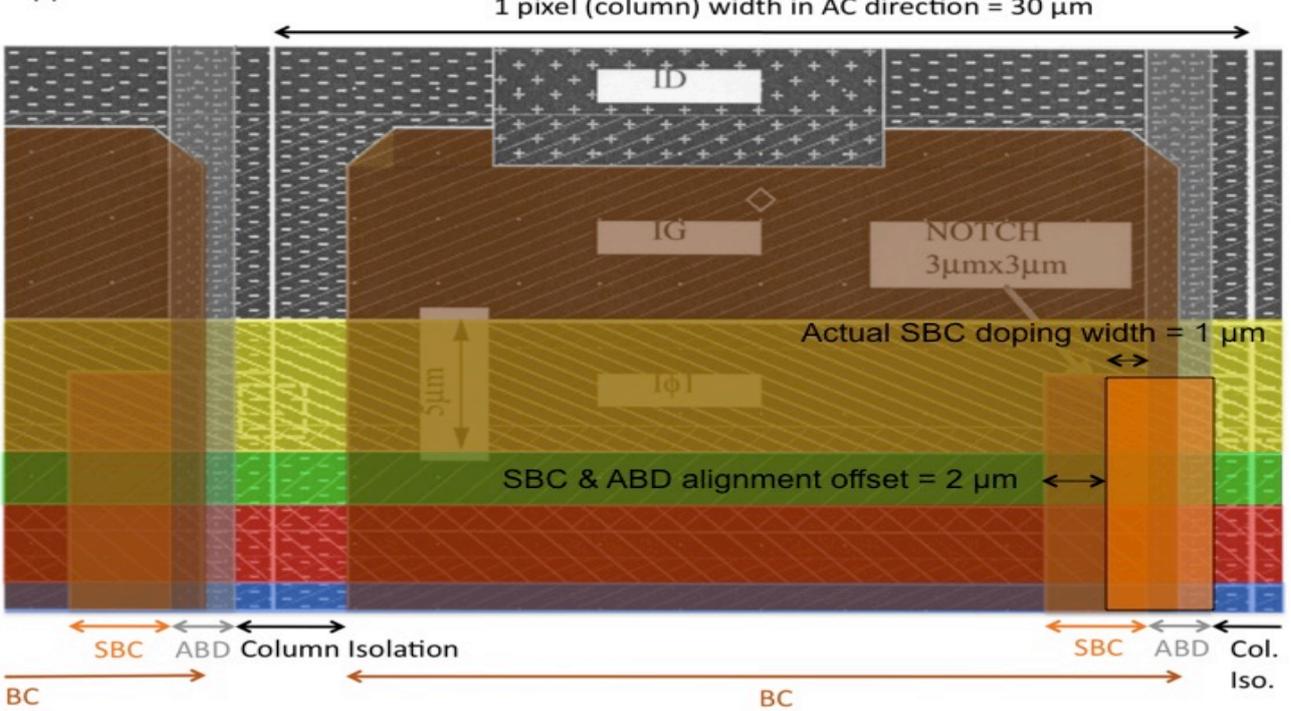

Fig. 6. Architecture of the Gaia e2v CCD91-72 pixel. *Top*: e2v's nominal design. *Bottom*: SBC and ABD alignment offset inferred from the simulated effective SBC doping width that has the capacity that matches measurements in the upper half of the tested CCD.

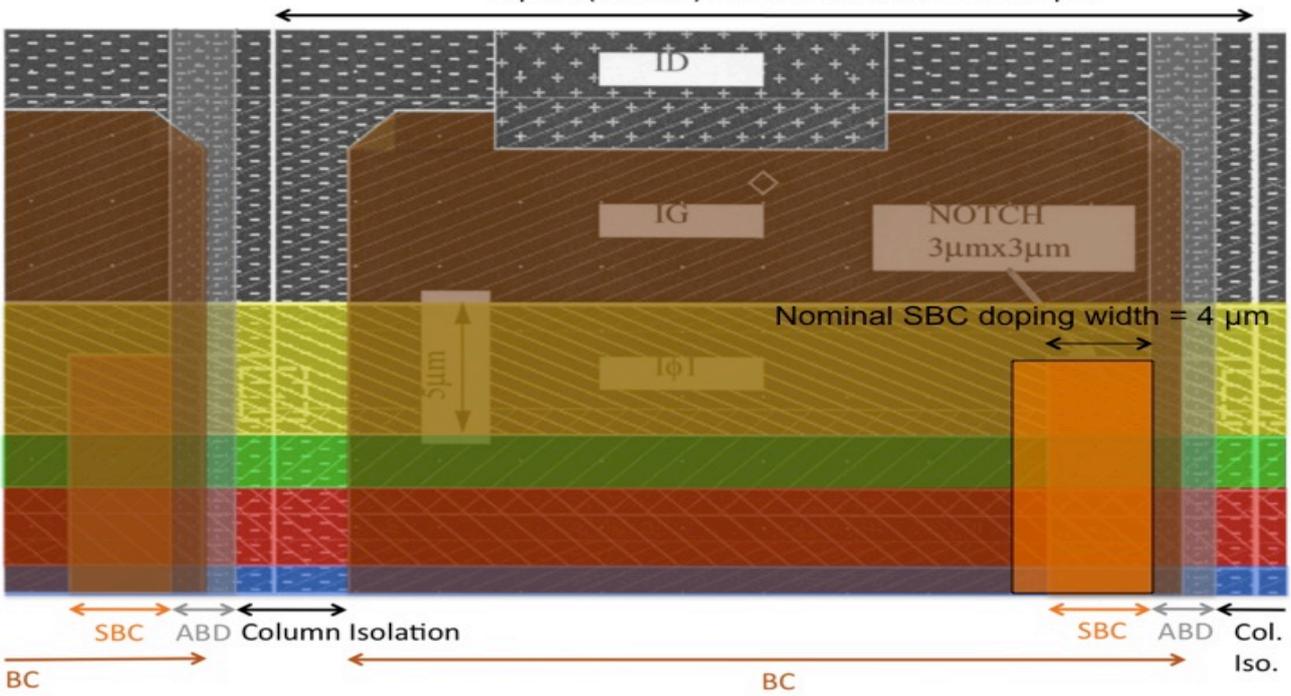
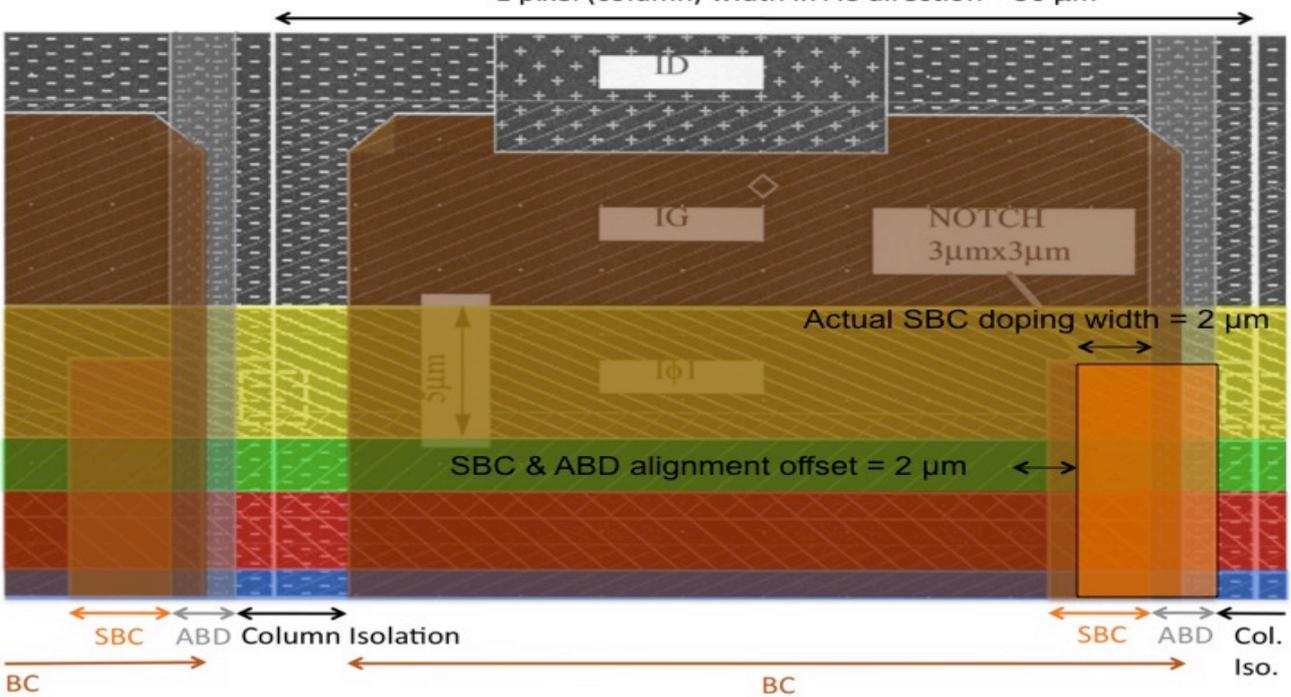

Fig. 7. Architecture of the Gaia e2v CCD91-72 pixel. *Top*: e2v's nominal design. *Bottom*: SBC and ABD alignment offset inferred from the simulated effective SBC doping width that has the capacity that matches measurements in the lower half of the tested CCD.

The alignment procedure occurs independently in each of the 18 stitch blocks in each Gaia CCD. This accounts for the spread of measured SBC capacities in Table 1. Applying e2v's predicted worst-case stitch error (-0.5 μm) to the nominal SBC doping widths of 3 and 4 μm, gives 2.5 and 3.5 μm wide SBCs, which, when simulated in the pixel model, Fig. 5 shows still yield capacities much larger than measured. Applying e2v's predicted worst-case stitch error (±0.5 μm) to the simulated SBC doping widths that yield that same capacities as measured, gives 1.5 and 2.5 μm wide SBCs in the lower half of the CCD, which, when simulated in the pixel model, Fig. 5 shows yield capacities much smaller and larger than measured.

In the upper half of the CCD, the simulated SBC doping widths that yield that same capacities as measured are only constrained to an upper limit of 1.5 μm. Applying e2v's predicted worst-case stitch error (+0.5 μm), gives 2 μm wide SBCs, which Fig. 5 shows yields a capacity much smaller than measured in the upper half of the CCD. This could either be explained by a SBC doping width of 1.5 μm with stitch errors <0.5 μm or a narrower SBC. Applying half e2v's predicted worst-case stitch error (±0.25 μm) to a 2 μm wide SBC, gives 1.75 and 2.25 μm wide SBCs, which, when simulated in the pixel model, Fig. 5 shows yields a range of capacities in good agreement with the range of measured capacities. If we consider the 2 μm positional offset between ABDs and SBCs to be a systematic stitch offset, the previous sentence suggests that the worst-case random stitch error between two pixel features within a stitch block is actually ±0.25 μm. Kohley et al. (2009)'s fig. 9[3] shows that the majority of stitch blocks in the lower half have SBC capacities of ~2000 electrons, corresponding to random stitch errors <0.25 μm. Fig. 5 indicates the simulation matched to the measurements is the one with a SBC doping width of 1 μm. This conclusion was originally based on relative differences being better controlled by e2v than absolute differences, meaning the 1 μm doping width difference between the SBCs in the two halves of the CCD is more likely to be preserved, giving SBC doping widths of 1 and 2, instead of 3 and 4, μm, than the absolute differences in their doping widths (1/2 versus 3/4).

However, Seabroke et al. (2010b)[6] will present preliminary evidence that the majority of Gaia CCDs made in 2003 do not preserve the 1 μm doping width difference between the SBCs in the two halves of the CCD. Instead, modelling of Gaia's Time-Delayed Integration suggests most 2003 CCDs have the same SBC capacities in both CCD halves (~1000-4000 electrons), which can only be explained with the same SBC doping widths in both halves. This requires two systematic offsets in the 2003 CCDs: ~1 μm in the upper halves and ~2 μm in the lower halves. With a sample size of one 2005 CCD (Kohley et al. 2009[3]), more need to be tested to establish if the majority only have a single systematic offset. Neil Murray (e2v centre for electronic imaging) plans to repeat Kohley et al. 2009[3]'s testing on a 2006 Gaia test structure. If this device is like Kohley et al. 2009[3]'s, it would add more weight to the possibility that the reason for the difference between 2003 and 2005/2006 CCDs could be that are e2v changed their lithographic mask set between these batches. The mask sets should be exactly the same so this reason can only be considered as circumstantial evidence for the moment.

Our claims of systematic offsets in e2v photolithography, which could be due to stitch offsets or lateral ABD shield doping diffusion, are based on our pixel model so it is natural to question the validity of this model. Due to the overlapping doping features and a lack of knowledge of the realistic doping profiles (except for BC, see Seabroke et al. 2009[2]), the model uses step doping profiles for simplicity. This means the resulting potential profiles are an approximation to reality. While the pixel FWC using step and realistic doping profiles should be checked that they are the same (this can only be done with the BC), because the potential profiles are similar, the FWC (and thus SBC capacity) should also be similar.

e2v simulations of an isolated ABD suggest its depth is not important but the lateral diffusion of the shielding is key. Two different e2v simulations agree the nominal lateral diffusion of the shielding is 1 μm, which is also included in our pixel model. The shielding has a gradient, which we have modelled with a step doping profile, using the doping levels which we derived in Section 2.3 according to Gaia operational constraints. They agree with the doping gradient derived by e2v modelling. Therefore, our pixel model is a simplification of the e2v design but simultaneously matches every measurable constraint, giving confidence in our model predictions.

# 4. CONCLUSIONS

We have presented our 3D Silvaco ATLAS model of the Gaia e2v CCD91-72 pixel (Astrometric Field (AF) variant). We reiterate our conclusions from Seabroke et al. 2009[2]: "This is the first attempt by anyone to model the Gaia pixel [as a whole] to derive its electron distributions. It may also be the first time this has been done for any e2v pixel. If this is not the first attempt by anyone to model an e2v pixel, it is the most complex e2v pixel to be modelled, as the Gaia image pixel is the most complex pixel architecture ever made by e2v, because it includes so many different features: buried channel (BC), supplementary buried channel (SBC) and anti-blooming drain (ABD) and shield."

Kohley et al. (2009)[3] measured the SBC capacities of a close-reject Flight Model (FM) AF CCD. They found that there were no working SBCs in the upper half of their CCD but the SBCs in the lower half had capacities ranging from 1000-3000 electrons. Using the pocket pumping technique, this was the first time that SBC capacities in the two halves of a Gaia CCD had been measured. Previously, only average SBC capacities of ~1000 electrons over the whole CCD have been estimated from First Pixel Response (FPR) curves[7,8]. 1000 electrons has been assumed to be the design goal for the storage capacity of the SBC. However, SBC capacities have never been a formally agreed CCD acceptance criterion in the Gaia contract between ESA/EADS Astrium and e2v. This is may be why the e2v's design model predictions have never been published before being included in this paper, as a result of this work's investigations to explain Kohley et al. (2009)[3]'s unexpected results. Moreover, e2v would not have agreed to such small SBC capacities as the doping width required is too small to guarantee a minimum capacity. Indeed, e2v argued for wider doping widths (4 and 5 $\mu$m) when the pixel was being designed.

Having previously successfully benchmarked the Silvaco software in Seabroke et al. (2009)[2], we have shown in this paper that our Silvaco pixel model is physically realistic. The main result in this paper is that our pixel model cannot explain Kohley et al. (2009)[3]'s SBC capacities using the nominal e2v design values for the SBC doping widths. (Using these values our model agrees with e2v's nominal SBC capacities, which are an order of magnitude greater than those measured.) We have found the SBC doping widths that yield the measured SBC capacities. These widths are systematically 2 $\mu$m smaller than the nominal widths in both CCD halves. Worst-case random alignment stitch errors (-0.5 $\mu$m) applied to the nominal SBC doping widths cannot explain the offsets. The range of SBC capacities in the lower half of the CCD were used to derive the worst-case random stitch error between two pixel features within a stitch block in this CCD to be ±0.25 $\mu$m. Therefore these offsets appear to be uncalibrated systematic offsets in e2v photolithography, which could either be due to systematic stitch offsets or lateral ABD shield doping diffusion. Our pixel model does not specifically simulate either of these scenarios, rather it simulates the same effective SBC doping width that is produced by both of these scenarios.

Our pixel model cannot constrain the manufacturing reason for the systematic offsets because the doping levels are derived from electrical measurements, rather than simulating e2v's photolithography processes, which would require a very detailed understanding of e2v's manufacturing. Nevertheless our results provide constraints for an investigation into e2v's processes. For example, if the 2 $\mu$m offset is due to stitch alignments and the offset occurred in the other direction, away from the ABD, then there would not be any overlap of the SBC doping with the shielding doping and so the resulting SBC capacities in both halves of the detector would be the nominal e2v design values.

Assuming the systematic offset is due to stitch alignments and the lateral ABD shield doping diffusion is well controlled, the sensitivity of the capacity of a SBC (adjacent to an ABD) to the alignment could be used as an independent measure of the alignment accuracy, via pixel modelling. This could be useful for Gaia (see below) and other high-precision CCD applications.

**4.1 Implications for Gaia science: radiation damage**

Astrium is currently conducting testing on radiation damaged Gaia CCDs. Seabroke et al. (2010b)[6] will present preliminary evidence that the CCDs analysed thus far have the same SBC capacities in both halves (~1000-4000 electrons), which can only be explained with the same effective SBC doping widths in both halves. As expected, Seabroke et al. (2010b)[6] will also show that CCDs without SBCs in their upper halves suffer worse charge loss than CCDs with SBCs in both halves. The best case scenario in terms of radiation damage is that the CCD measured by

Kohley et al. (2009)[3] is an single, isolated, extreme outlier in terms of e2v manufacturing spreads and that all FM CCDs have working SBCs in their upper halves. The worst scenario is that all FM CCDs are like the one tested by Kohley et al. (2009)[3]. This would mean the radiation damage test results are not representative of FM CCDs and so its analysis presents a too-optimistic view on the effect of radiation damage on Gaia data. Alternatively, there could be a mixture of the two scenarios. Our pixel model cannot constrain the manufacturing reason for the range of SBC capacities, so it does not allow us to predict which scenario is most likely to reflect reality.

Even if Gaia data will suffer more radiation damage than currently predicted, the science will not suffer if the model for correcting radiation damage is sufficiently realistic. The current model[9] may need to be improved in light of future test results. Seabroke, Holland & Cropper (2008)[1]'s fig. 6 shows even the brightest stars Gaia will observe spend some time accumulating charge in SBCs. However, for the majority of the CCD exposure time the charge packets of these stars within the image point spread function are larger than the SBC capacity so the electrons fill the SBCs and consequently collapse the SBCs' potentials into the BCs' potentials. Seabroke, Holland & Cropper (2008)[1]'s fig. 6 also shows the charge packets of the more numerous faintest stars in the AF and Blue and Red Photometer CCDs and all the stars in the Radial Velocity Spectrometer (RVS) CCDs spend the entire exposure time in SBCs, rather than BCs. All this suggests a better understanding of the FM SBCs can only be beneficial for developing the Gaia radiation damage correction model.

**4.2 Implications for Gaia science: geometric calibration models**

If the systematic offsets are caused by lateral ABD shield doping diffusion, then there is no reason to suspect the stitch errors are not nominal. However, if the systematic offsets are caused by systematic stitch offsets, then this could yield a much larger than currently expected geometric distortion:

"The probable worst-case stitch error in the column isolation pattern in the across- scan X direction would give no error in pixel dimension or pitch, but an across-scan displacement of 0.25 μm of one sub-array relative to its theoretical position. The only component of the stitch error which could be an accumulating error is the formation of the "zero grid", all subsequent layers being subject to random alignment errors relative to that grid. The image area of the Gaia AF device contains 8 stitch boundaries … so the worst-case error in the across-scan X direction positioning due to "zero grid" errors would be $8 \times 0.25 = 2$ μm. It is improbable that these tolerances would all add cumulatively and a figure of 0.7 μm (found from the square root of the sum of squares assumption for uncorrelated error sources) is believed to be more representative."[5] "The use of the Nikon for the zero-grid layer should have reduced the potential accumulating error from 0.7 μm to ~0.2 μm."[10]

Applying this to a systematic stitch error of 2 μm gives $8 \times 2 = 16$ μm for the worst-case error but the more realistic error should be $< \sqrt{8 \times 2^2} \sim 6$ μm. "The averaged pixel locations will be routinely calibrated (both AL [Along Scan (charge transfer direction)] and AC [Across Scan]) in AGIS [Astrometric Global Iterative Solution] – and partly in ODAS [One Day Astrometric Solution] – in the form of small-scale geometric calibration coefficients. … The stitch boundaries will create large discontinuities in the geometry which must be modeled appropriately."[10]

Therefore, if the systematic offsets are caused by the systematic stitch offsets, then, as previously mentioned, using our pixel model to interpret SBC capacities, in terms of stitch alignment accuracy could be the only method to verify AGIS and ODAS geometric residuals. There are currently no plans to determine FM SBC capacities before launch but if need be they could be constrained in-flight using different charge injection levels to populate FPR curves, which can be interpreted in terms of SBC capacities with modelling (see Seabroke et al. 2010b[6] for more details).

# ACKNOWLEDGMENTS

The Silvaco ATLAS software license fee is generously funded by the UK VEGA Gaia Data Flow System grant thanks to F. van Leeuwen and G. Gilmore (Institute of Astronomy, Cambridge). Thanks also go to the members of the Gaia Radiation Task Force for stimulating meetings and discussions.